\documentclass[12pt]{article}
\usepackage[english]{babel}
\usepackage{amsfonts}
\usepackage{amsmath}
\usepackage{amssymb}
\usepackage{mathrsfs}
\usepackage{latexsym}
\usepackage{multicol}
\usepackage{fancybox} 
\usepackage{color}
\usepackage{hyperref}
\usepackage{graphicx} 
\usepackage{caption}
\usepackage{enumerate}
\usepackage{upgreek}
\usepackage[overload]{empheq}

\def\be{\begin{equation}}
\def\ee{\end{equation}}

\def\d'{``}

\def\be{\begin{equation}}
\def\ee{\end{equation}}
\def\bea{\begin{eqnarray}}
\def\eea{\end{eqnarray}}

\def\i'{\textrm{i}}

\def\d'{``}

\begin{document}

\title{Some numerical observations about the COVID-19 epidemic in Italy} 
\author{Federico Zullo\thanks{DICATAM,  Universit\`a degli Studi di Brescia, Brescia, Italy}}





\maketitle 

\begin{abstract}
\noindent
We give some numerical observations on the total number of infected by the SARS-CoV-2 in Italy. The analysis is based on a tanh formula involving two parameters.  A polynomial correlation between the parameters gives an upper bound for the time of the peak of new infected. A numerical indicator of the temporal variability of the upper bound is introduced. The result and the possibility to extend the analysis to other countries are discussed in the conclusions. 
\end{abstract} 

\section{Introduction} \label{intro}
\vspace*{4mm}

In most epidemics it is impossible to determine the true number of new infected individuals per day.  This is the case for the new coronavirus disease, since  asymptomatic people or with very mild symptoms may not seek medical assistance and cannot be identified \cite{Baud}. Realistic data are fundamental to understand the epidemic and to steer the efforts to inhibit the disease in the right direction.  Also, the dynamical variables of epidemiological models usually are linked to, or describe directly, the evolution of the true number of infected: the comparison with the empirical data may be problematic if those numbers are not realistic. On this side, researches about  the estimation of the real scale of the epidemic or of the proportion of the asymptomatic already appeared in the literature (see e.g. \cite{Lietal} or \cite{Kenji}).

On the other hand, under very reasonable hypotheses, it is possible to assume that suitable measurable quantities are determined by the relative values of certain characteristics of the population only (in opposition to global absolute values): in this case the knowledge of only a fraction of new infected individuals per day may still be useful to estimate some of the measurable quantities. This property (we will refer to it as ``scale invariance'') must be reflected in a scale-independent property of the underlying epidemiological model.  In this paper we assume that the time of the peak of new infected by the SARS-CoV-2 in Italy has the scale-invariance property. We are aware of the fact that this assumption can be considered at best a rather crude approximation to a very complex system of interactions. In our opinion however, when taken as a working hypothesis, it can provide a well-founded basis, or at least a staring point, to achieve reasonable estimates. This point of view will be justified further in section (\ref{sec1}) on the basis of the SIR epidemiological model. 

Since the start of the epidemic in China, a certain number of studies appeared in the mathematical community about this subject: the description of the spatial or temporal diffusion of the infected in given regions \cite{FG}, \cite{Gaeta1}-\cite{Giu}, the transmission dynamics of the infection \cite{K}, the economic and financial consequences of the epidemic \cite{Cal}, the effect of atmospheric indicators on the spread of the virus \cite{J}, are only  a fraction of the topics under investigation in these days. A certain number of epidemiological studies are connected to the SIR model. The SIR model is one of the simplest non-linear deterministic continuous (in time) model of epidemiology: the overall population is divided in three disjoint classes: $S$, i.e. the number of susceptible individuals, $I$, the number of infectious individuals and $R$, the number of recovered individuals. Albeit its non-linearity, the dynamic of the model is fairly uncomplicated and manageable from an analytical point of view and displays very interesting and realistic properties such as the existence of an \emph{epidemic threshold} (see e.g. \cite{Braun} \cite{Murray}). 

We must underline that the assumption of the scale invariance is not subjected to the adherence of the empirical data to this particular mathematical model. The SIR model is seen here as an instance of the family of models possessing the scale invariance.

The paper is organized as follows: in section \ref{sec1} the SIR model is introduced and briefly discussed. In section (\ref{sec3}) we  analyze the data of the cumulative number of infected in Italy on the base of two simple hypotheses. An upper bound for the timing of the peak of new number of infected is obtained. This upper bound is dynamic: when more data are added to the model in the course of the epidemic it may changes in time. In section (\ref{sec4}) we will discuss the predictive validity of the model on the basis of  a numerical indicator measuring the temporal variability of the upper bound.  In the conclusions we will comment about the results and look for possible extensions.

\section{The SIR model and the scale invariance property}\label{sec1}
\vspace*{4mm}
The SIR model describes the evolution of the individuals in the  susceptible, infectious and recovered classes with the following differential equations:
\begin{equation}\label{SIR}\left\{\begin{split}
& \frac{dS}{dt}=-r\frac{SI}{N},\\
&\frac{dI}{dt}=r\frac{SI}{N}-aI,\\
&\frac{dR}{dt}=aI.
\end{split}\right.\end{equation}
The total population $N=S+R+I$ is a conserved quantity from the dynamical point of view, meaning that there are only two independent variables in the set of equations (\ref{SIR}). The characteristics of this model are well-known and the interested readers can look for example at the discussions in the classical books of Braun \cite{Braun} and Murrray \cite{Murray}. Here we will make only few observations, relevant for the next sections.

Some authors do not include the denominator $N$ on the right hand side of (\ref{SIR}), since it is a constant and can be absorbed by a re-definition of the parameter $r$. However, we will keep it: in this way it is indeed evident  the scale invariance property of the model: if the initial conditions $(S_0,I_0,R_0)$ are scaled by a common constant factor $\alpha$, (and so the total population is scaled by a factor $\alpha$), the solution is scaled by the same factor. Indeed it is enough to observe that, if $(S(t),I(t),R(t))$ are the solutions of equations (\ref{SIR}) corresponding to the initial conditions $(S_0,I_0,R_0)$, then $(\alpha S(t),\alpha I(t),\alpha R(t))$ are the solutions corresponding to the initial conditions $(\alpha S_0,\alpha I_0,\alpha R_0)$.

Some temporal properties of this model, like the time corresponding to a maximum in $I$ (the time of the peak of the infected), do not depend on the scaling. This property is very useful, since the actual number of infected or susceptible (and then of recovered) is in general not known. The reasonable assumption that the \emph{same fraction} (with respect to the total) of infected, susceptible and recovered individuals are known, gives  the possibility, in this case,  to compare the measured data with the properties that are scale-independent.    

The solution of the system (\ref{SIR}) can be written in terms of just one variable: if $R(t)$ is known, $I(t)$ can be obtained by the third equation and $S(t)$ from the first one or from the constrain $N=S(t)+I(t)+R(t)$. If the epidemic is not severe (the number $R(t)$ can be considered small compared to the overall population), an explicit formula for the number of recovered can be obtained in terms of the hyperbolic tangent function. The functions reads as
\begin{equation}\label{R}
R(t)=\alpha \tanh(\beta t -c)+\alpha \tanh(c)
\end{equation}
where we used the initial value $R(0)=0$ and the parameters $(\alpha, \beta, c)$ can be made explicit in terms of the parameters appearing in (\ref{SIR}). The previous is one of the example of the so-called s-shaped epidemiological curve (with a ``peaked'' derivative, the function sech$^2$) that universally describes an infection disease.   The value of the number of infected can be obtained by derivation, i.e. $I(t)=\alpha\beta\textrm{sech}(\beta t-c)^2$. When considering the cumulative number of infected, $R+I$, the contribution of sech$^2$ is negligible on the tails, whereas it is more pronounced in correspondence of the maximum of sech$^2$, but it is however small if the value of the parameter $\beta$ is less than one. In this case, the value of $R+I$ is well approximated by a $\tanh$ formula like (\ref{R}), with a certain different value of $c$. This is what we will assume in the next section.

\section{Analysis of data with a $\tanh$ model}\label{sec3}
The discussion made at the end of the previous section, despite to be very basic, has the advantage to be manageable and to incorporate the main properties of the SIR model. It is not by chance that the first application of the SIR model  (the Bombay plague of 1905) by  Kermack and McKendrick \cite{KM} used precisely the $\tanh$ formula above. \\
In the following we will base our analysis on two hypothesis:
\begin{enumerate}[(i)]
\item We assume that the cumulative number of infected is described by a $\tanh$ model. Although this assumption is coherent with the finding of the SIR model, it does not depend on the particular dynamical model considered,
\item We assume that, whatever it is the underlying model describing the evolution of the number of infected,  this model is scale invariant, in the sense specified in the previous section.
 \end{enumerate}
The second hypothesis is fundamental since we are going to look at  scale-independent quantities: even in the case the measured number of infected and recovered individuals are different from the actual values, it is possible to estimate these quantities.

The cumulative total number of infected that will be considered in the next lines are those of entire Italy territory. There are at least two reasons that suggested to not take regional or local data: the first one is that the epidemic started to spread across three different regions (Lombardy, Veneto and Emilia-Romagna) and there could not be a correspondence between the locality where a certain fraction of inhabitants reside and the region where this fraction was infected. This is also true at a national level, but the fraction is assumed to be smaller. The second reason is that a non negligible number of workers and students moved, just before the lockdown, from the regions in the north of Italy to their regions of origin in the center and south of Italy. The possibility that a non negligible flow of infected people passed from the north to other regions  should be taken into consideration. By taking the entire national set of data we overpass the above issues.  

The data can be taken for example from Italian Protezione Civile \cite{PC}, from WHO \cite{W}, or from Worldometer \cite{WM}. The cumulative total number of infected will be indicated by $F_n$, with $F_1=21$ corresponding to the number of infected on 21st of February 2020. The subscript $n$ stays for the number of days from the starting of epidemic. These data will be opposed to  the continuous formula
\begin{equation}\label{f}
f(t,\alpha,\beta,c)=\alpha \tanh(\beta t-c)+\alpha \tanh(c).
\end{equation}
The value of $\beta$ will be taken to be constrained by the equation
\begin{equation}\label{beta}
\alpha \tanh(\beta -c)+\alpha \tanh(c)=F_1.
\end{equation}
The function $f$ (\ref{f}) then depends on two parameters, $\alpha$ and $c$. When necessary, to stress the dependence on these parameters, we will denote the function with $f_{\alpha,c}(t)$.  The cumulative final number of infected expected from  formula (\ref{f}) is given by $f_{\infty}=\alpha(1+\tanh(c))$. It is possible to estimate the parameters $\alpha$ and $c$ by minimizing the difference between the actual and predicted number of cases, i.e. minimizing
\begin{equation}
S_n=\sum_{i=2}^n \left(F_n-f(n)\right)^2
\end{equation}
In order to have a reasonable minimum number of data, we start the analysis by taking $n\geq 15$. The values of the parameters minimizing the sum $S_n$ are reported in table (\ref{t1})\footnote{Some values are slightly different from those of the first preprint version of the paper since some data on the cumulative number of infected $F_n$ have been corrected according to \cite{WM}.}

\begin{figure}
\centering
\includegraphics[scale=0.8]{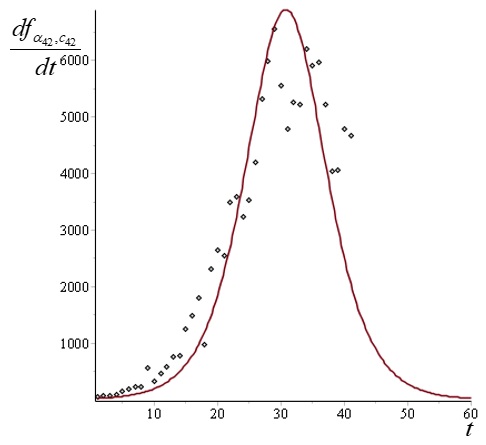}
\caption{The new number of infected and the continuous curve given by $df_{\alpha_{42},c_{42}}/dt$}
\label{fig1}
\end{figure}

\begin{figure}
\centering
\includegraphics[scale=0.8]{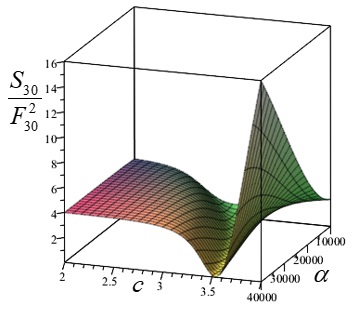}
\caption{The basin of depressed values of $S_{30}$: the values have been re-scaled to $F_{30}^2$ for easy of plotting}
\label{fig2}
\end{figure}

\begin{figure}
\centering
\includegraphics[scale=0.7]{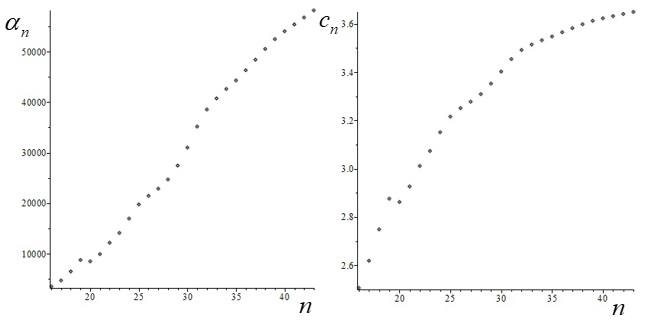}
\caption{The values of  $\alpha_n$ and $c_n$ vs $n$ as given in table (\ref{t1})}
\label{fig3}
\end{figure}

\begin{figure}
\centering
\includegraphics[scale=0.7]{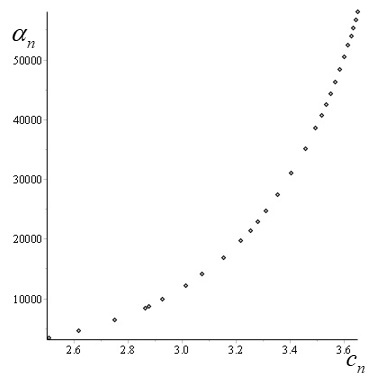}
\caption{The values of  $\alpha_n$ vs the values of $c_n$ as given in table (\ref{t1})}
\label{fig4}
\end{figure}

\begin{table}
\centering
 \begin{tabular}{||c c c c c c||}
 \hline
$ n$ & $\alpha_n$ & $c_n$ & $n$ & $\alpha_n$ & $c_n$\\ [0.5ex] 
 \hline\hline
 15 & 3514.0 & 2.506 & 29 & 30995.0 & 3.403 \\ 
 \hline
 16 & 4703.0 & 2.618 & 30 & 35166.7& 3.455 \\
 \hline
 17 & 6482.9& 2.749 & 31 & 38541.2 &3.492\\
 \hline
 18 & 8757.1 & 2.876 & 32 & 40737.4 & 3.515 \\
 \hline
 19 & 8748.9 & 2.862 & 33 & 42602.6 & 3.533\\ 
  \hline
 20 &9908.2 & 2.926 & 34 & 44297.6 & 3.548\\
  \hline
 21 & 12158.0 & 3.011 & 35  & 46317.8  & 3.566 \\
  \hline
 22 & 14102.9 & 3.074 & 36& 48397.7 & 3.582\\
   \hline
 23 & 16946.5 & 3.151 & 37 &50527.0 &3.599\\
   \hline
 24 & 19752.4 & 3.216 & 38 & 52472.3 &3.613\\
   \hline
 25 & 21453.2 & 3.251 & 39 & 54030.3 &3.624\\
   \hline
 26 & 22862.5 & 3.278 & 40 & 55380.4 & 3.633\\
   \hline
 27 & 24679.2 & 3.309 & 41 & 56736.1 & 3.642\\
   \hline
 28 & 2743.8 & 3.353 & 42 &58161.2 & 3.650
  
\end{tabular}
\caption{the estimated values of $\alpha_n$ and $c_n$ } \label{t1}
\end{table}
 A plot of $df_{\alpha_{42},c_{42}}/dt$ and of $F_{n+1}-F_n$ is reported in figure (\ref{fig1}).
A fundamental observation is that the function $S_n$ actually has a basin of depressed values, showed in detail in figure (\ref{fig2}) for  a given value of $n$. This basin of minimum seems to indicate that there is a given function $\alpha(c)$ giving a family of $\tanh$ curves with reduced values of $S_n$. The curve $\alpha(c)$ is quite stable by varying $n$ (see section (\ref{sec4})) and suggests to look at the values of $\alpha_n$ as functions of $c_n$.
In figure (\ref{fig3}) we report the plot of the values of $\alpha_n$ and $c_n$ given in table (\ref{t1}) as a function  of $n$, whereas in figure (\ref{fig4}) the plot of the values $(c_n,\alpha_n)$.  The values of $\alpha_n$ vs $c_n$, as explained above,  describe the basin of depressed values for $S_n$ as a function of $\alpha(c)$. We make a cubic fit, with linear coefficients, in order to get a rough description  of the curve $\alpha(c)$:
\begin{equation}\label{ac}
\alpha= \sum_{k=0}^3 a_kc^k
\end{equation}
 Clearly, by considering a number $N$ of values of $\alpha_n$ and $c_n$ to fit $a_k$, k=0,...,3, we will obtain a set of values $\{a_{k,N}\}$. By fitting all the data available (i.e. by taking $N=28$), we get the following values for the coefficients $a_k$:
 \begin{equation}\label{eqres}
 a_0=-999707,\;  a_1=1077192 ,\; a_2=-389358,\; a_3=47555.\;
  \end{equation}
 It is possible to get more terms in the sum (\ref{ac}), but the cubic term is sufficient to get a formula accurate  enough to what we are going to say. 

\begin{figure}
\centering
\includegraphics[scale=0.7]{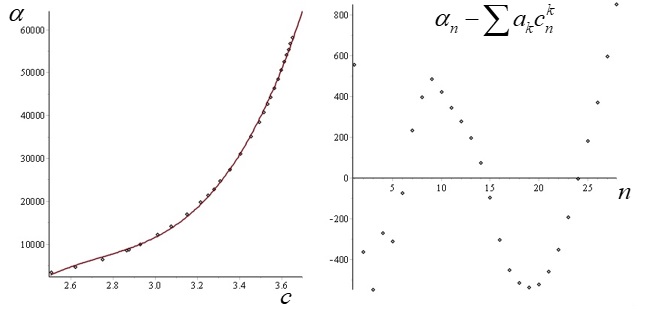}
\caption{The plot of the fit (\ref{ac}) (left) and the values of the residuals $\alpha_n-\sum_{k=0}^3 a_kc_n^k$ (right)}
\label{fig5}
\end{figure}

\begin{figure}
\centering
\includegraphics[scale=0.7]{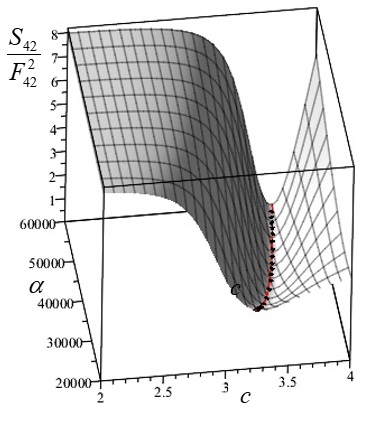}
\caption{The basin of minima together with the curve  (\ref{ac}) (in red)  and the actual values $(c_n,\alpha_n)$ (black dots)}
\label{fig6}
\end{figure}

The plot of the fit is given in figure (\ref{fig5}), together with the values of the residuals $\alpha_n-\sum_{k=0}^3 a_kc_n^k$, where the values $a_k$ are those given in equation (\ref{eqres}). A comparison between the curve $\alpha(c)$ and the basin of minima for $S_n$ has been plotted in figure (\ref{fig6}): the red curve is the function (\ref{ac}) with the black dots giving the actual values of $(c_n,\alpha_n)$ in table (\ref{t1}).

The function $\alpha(c)$ denotes a trend in the data that may be useful. If in the next days the values of the infection continue to rise, it is reasonable to expect that the values of $\alpha$ and $c$ will be constrained closely by the same curve. Clearly the model used here is rough, but it can give at least an idea about the future trend of the data. We are tacitly assuming that there will be no other cluster of infection around Italy in the next days: the point will be discussed later.

\begin{figure}
\centering
\includegraphics[scale=0.8]{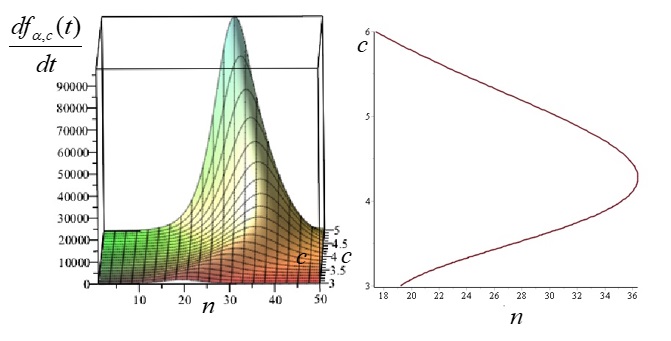}
\caption{The values of the derivative of the function  $f(t)$ vs $n$ and $c$ (left) and the corresponding values of maxima as a function of $n$ and $c$ (the points where the second derivative of $f(t)$ vanishes) (right).}
\label{fig7}
\end{figure}

Now we consider the function $f$ in (\ref{f}) as  a function of $t$ and $c$ alone, since the value of $a$ is constrained by the curve  (\ref{ac}). The plot of the derivative of this function (with respect to $t$) gives the time of the peak of infections. The plot is reported in figure (\ref{fig7}): we notice that the maximum of the derivative of the cumulative number of infected increases with $n$ up to $c\sim 4.3$ and then \emph{decreases} by increasing $c$. This gives an upper bound for the peak of new number of infected (the point where the second derivative of $f(t)$ (\ref{f}) is zero), given by 36 days after the data corresponding to  $F_1$ (21st of February). 

\section{Temporal variability of the upper bound}\label{sec4}
\begin{figure}
\centering
\includegraphics[scale=0.7]{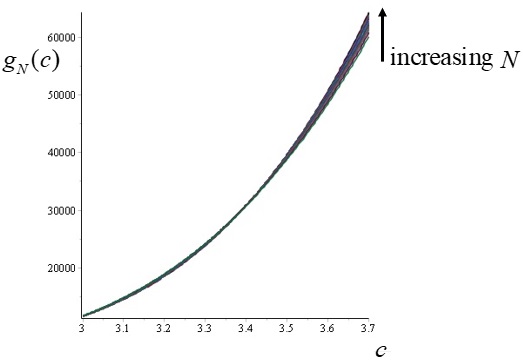}
\caption{The functions $g_{N}(c)$, for $N=30...42$: the curves overlap in a small region of the plane $(\alpha,c)$.}
\label{fig8}
\end{figure}

\begin{figure}
\centering
\includegraphics[scale=0.7]{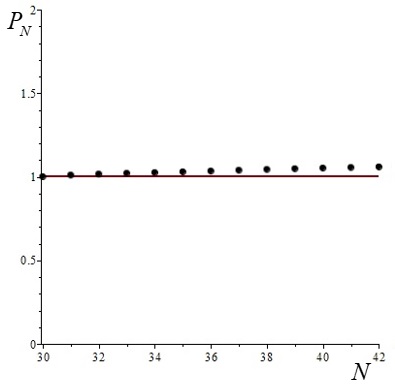}
\caption{The values of the parameters $P_N$, $N=30...42$, measuring the temporal variability of the curves $g_N(c)$.}
\label{fig9}
\end{figure}

The basin of minimum of $S_n$, for each fixed $n$, is described by a function $g_n=\alpha(c)$: this function gives a family of $\tanh$ curves with reduced values of $S_n$. The curve $\alpha(c)$ has been described in the previous section by fitting the values of $\alpha_n$ and $c_n$ in table (\ref{t1}) with a cubic formula. We obtained just one curve by making use of all the data available, i.e. 28 couples $(c_n,\alpha_n)$. It is possible to ask how the curves $g_n(c)$ depend on the number of the data available: if the curves $g_n$ have a temporal stability, then they can be used to make a reliable estimation of the time of the peak of new infected. To address this question we fit the data $(c_i,\alpha_i)$, with the index $i$ from 15 to $N$, by varying the number of data taken (i.e. by varying $N$). The fit is again cubic, like in (\ref{ac}). In this way we follow the temporal variation of the curve $g_N(c)$. In plot (\ref{fig8}) we report the curves $g_N(c)$ for $N$ in the interval $[30,42]$: these are the last thirteen curves (in order of time). 
It is possible to see that indeed the convexity of the curves is slowly increasing, so that the differences are more pronounced for higher values of $c$. To have a measure of the variability of these curves we introduce a parameter giving the relative increase of $g_n(c)$ for $c$ fixed and equal to the last available value (i.e. $c=c_{42} in table (\ref{t1}))$ 
\begin{equation}\label{par}
P_N= \left.\frac{\sum_{k=0}^3 a_{N,k}c^k}{\sum_{k=0}^3 a_{30,k}c^k}\right|_{c=c_{42}}
\end{equation}
 The values of $P_N$, $N=30...42$, are also reported in figure (\ref{fig9}).


\section{Conclusions}
The above analysis, despite using a rough function for the total number of infected,  is able to give an upper bound for the time of the peak of new infected (27th of March) thanks to the observation that the values of $\alpha_n$ are, in a certain sense, not independent on the values of $c_n$ and are well described by a polynomial interpolation with linear coefficient. The hypothesis about the scale invariance of the underlying model (that, we repeat, not necessarily is represented by the SIR model) and the low temporal variability of the upper bound are fundamental for the accuracy of the result. Another underlying assumption is that the restrictive measures will be kept and observed in the next days and there will be no other clusters in the south of Italy (in the SIR model language, the values of $S_0$ are below the epidemic threshold, see e.g. \cite{Murray}). In the unfortunate case that there will be other clusters, it is possible to think at a substitution of the $\tanh$ curve by a combination of such functions: if there are two clusters of comparable magnitude, then we will have
\begin{equation}\label{f2}
f(t)=\alpha_1 \tanh(\beta_1 t-c_1)+\alpha_1 \tanh(c_1)+\alpha_2 \tanh(\beta_2 t-c_2)+\alpha_2 \tanh(c_2)
\end{equation}  
In a next paper other sets of data, from different countries, will be analyzed.


\small 
\begin{thebibliography}{40}
\bibitem{Baud} D. Baud, X. Qi, K. Nielsen-Saines, D. Musso, L. Pomar, G. Favre: Real estimates of mortality following COVID-19 infection, Lancet, Infectious Diseases, 2020, DOI: https://doi.org/10.1016/S1473-3099(20)30195-X
\bibitem{Cal}  C. Albulescu: Coronavirus and oil price crash, https://arxiv.org/abs/2003.06184 
\bibitem{Braun} M. Braun: Differential Equations and Their Applications: An Introduction to Applied Mathematics, Springer, ISBN 978-0387978949.
\bibitem{Ga1} A. Lai, A. Bergna C. Acciarri, M. Galli, G. Zehender: Early phylogenetic estimate of the effective reproduction number of SARS‐CoV‐2, Journal of Medical Virology,  https://doi.org/10.1002/jmv.25723, 2020, see also the interview to Prof. Galli to the italian newspapaer Corriere della Sera https://www.corriere.it/cronache/20\_marzo\_01/galli-il-coronavirus-italia-settimane-tsunami-il-sistema-sanitario-a153160c-5b3d-11ea-8b1a-b76251361796.shtml
\bibitem{FG} D.  Fanelli,  F. Piazza: Analysis and forecast of COVID-19 spreading in China, Italy and France, https://arxiv.org/abs/2003.06031
\bibitem{J} Jingyuan Wang, Ke Tang, Kai Feng, Weifeng Lv: High Temperature and High Humidity Reduce the Transmission of COVID-19, https://arxiv.org/abs/2003.05003.
\bibitem{Kenji} K. Mizumoto, K. Kagaya, A. Zarebski, G. Chowell: Estimating the asymptomatic proportion of coronavirus disease 2019 (COVID-19) cases on board the Diamond Princess cruise ship, Yokohama, Japan, 2020, Euro Surveill. 2020 Mar 12; 25(10): 2000180. doi: 10.2807/1560-7917.ES.2020.25.10.2000180
\bibitem{K} Kucharski AJ, Russell TW, Diamond C, Liu Y, Edmunds J, Funk S, Eggo RM: Early dynamics of transmission and control of COVID-19: a mathematical modelling study, Lancet Infect Dis. 2020 Mar 11. pii: S1473-3099(20)30144-4. doi: 10.1016/S1473-3099(20)30144-4.
\bibitem{KM} Kermack, W. O.; McKendrick, A. G.: A Contribution to the Mathematical Theory of Epidemics, Proceedings of the Royal Society A. 115 (772): 700–721, 1927
\bibitem{Lietal} Dalin Li, Jun Lv, Gregory Botwin, Jonathan Braun, Weihua Cao, Liming Li, Dermot P.B. McGovern. Estimating the scale of COVID-19 Epidemic in the United States: Simulations Based on Air Traffic directly from Wuhan, China. submitted to MedRxiv, 2020 DOI: 10.1101/2020.03.06.20031880 
\bibitem{Gaeta1} G. Gaeta: Data analysis for the COVID-19 early dynamics in Northern Italy, https://arxiv.org/abs/2003.02062
\bibitem{Gaeta2} G. Gaeta: Data Analysis for the COVID-19 early dynamics in Northern Italy. The effect of first restrictive measures, https://arxiv.org/abs/2003.03775
\bibitem{Giu}  D. Giuliani, M. M. Dickson, G. Espa, F. Santi: Modelling and predicting the spatio-temporal spread of Coronavirus disease 2019 (COVID-19) in Italy, https://arxiv.org/abs/2003.06664
\bibitem{Murray} J. Murray: Mathematical Biology, Vol. 1, Springer, Berlin, 2002, ISBN 0-387-95223-3
\bibitem{PC} http://www.protezionecivile.gov.it/media-comunicazione/comunicati-stampa
\bibitem{W}  https://www.who.int/emergencies/diseases/novel-coronavirus-2019/situation-reports
\bibitem{WM} https://www.worldometers.info/


\end{thebibliography}
\end{document}